\documentclass[journal=acsnano,manuscript=article]{achemso}

\usepackage{chemformula} 
\usepackage[T1]{fontenc} 
\usepackage{bm}
\usepackage{amssymb}
\usepackage{xcolor}
\usepackage{ulem}
\usepackage{hyperref}

\author{Tristan Riccardi}
\affiliation[LNCMI]
{Laboratoire National des Champs Magnétiques Intenses, LNCMI-EMFL, CNRS UPR3228,Univ. Grenoble Alpes, Univ. Toulouse, Univ. Toulouse 3, INSA-T, Grenoble and Toulouse, France}
\alsoaffiliation[Néel]
{Univ. Grenoble Alpes, CNRS, Grenoble INP, Institut NEEL, 38000 Grenoble, France}

\author{Amit Pawbake}
\affiliation[LNCMI]
{Laboratoire National des Champs Magnétiques Intenses, LNCMI-EMFL, CNRS UPR3228,Univ. Grenoble Alpes, Univ. Toulouse, Univ. Toulouse 3, INSA-T, Grenoble and Toulouse, France}

\author{Shalini Badola}
\affiliation[LNCMI]
{Laboratoire National des Champs Magnétiques Intenses, LNCMI-EMFL, CNRS UPR3228,Univ. Grenoble Alpes, Univ. Toulouse, Univ. Toulouse 3, INSA-T, Grenoble and Toulouse, France}

\author{Floriant Petot}
\affiliation[Cinam]
{Aix-Marseille Universit\'e, CNRS, CINaM, Marseille, France}

\author{Beno\^it Gr\'emaud}
\affiliation{Aix Marseille Univ, Universit\'e de Toulon, CNRS, CPT, Marseille, France}

\author{Andres Saul}
\affiliation[Cinam]
{Aix-Marseille Universit\'e, CNRS, CINaM, Marseille, France}

\author{Kiran Singh}
\affiliation[LMM]
{Lab. of Multifunctional Materials, Department of Physics, Dr. B. R. Ambedkar National Institute of Technology, Jalandhar, Punjab 144008, India}

\author{Niranjana Renjith Nair}
\affiliation[Prague]
{Chemistry Department, University of Chemistry and Technology Prague, 16628 Prague, Czech Republic}

\author{Rasmiya Shirin Chemban}
\affiliation[Prague]
{Chemistry Department, University of Chemistry and Technology Prague, 16628 Prague, Czech Republic}

\author{Zdenek Sofer}
\affiliation[Prague]
{Chemistry Department, University of Chemistry and Technology Prague, 16628 Prague, Czech Republic}

\author{Johann Coraux}
\affiliation[Néel]
{Univ. Grenoble Alpes, CNRS, Grenoble INP, Institut NEEL, 38000 Grenoble, France}
\email{johann.coraux@neel.cnrs.fr}

\author{Clément Faugeras}
\affiliation[LNCMI]
{Laboratoire National des Champs Magnétiques Intenses, LNCMI-EMFL, CNRS UPR3228,Univ. Grenoble Alpes, Univ. Toulouse, Univ. Toulouse 3, INSA-T, Grenoble and Toulouse, France}
\email{clement.faugeras@lncmi.cnrs.fr}

\title[Title]
  {Magnetic field tuning of modulated magnetic orders in CrOCl at the two-dimensional limit}

\usepackage{graphicx}

\begin{document}

\begin{abstract}
Chromium oxychloride is a van der Waals magnet with intrinsic competing exchange interactions, including a strong antiferromagnetic one, source of a very rich magnetic phase diagram, with ferrimagnetic, antiferromagnetic, and canted states, up to high magnetic fields. We investigate the sequence of these magnetic phases in thin layers of CrOCl using magneto-Raman scattering spectroscopy. We identify phases whose magnetic order is commensurate with the atomic latice, and find signatures of strong magneto-striction, presumably of exchange origin. The coupling of the spin and atomic degrees of freedom in the crystal is observed down to the single-layer limit --- phonon modes significantly soften or stiffen, in a complex way due to the competition of interactions. The existence domains of the different phases change with the number of layers.
\end{abstract}

\newpage

\section{Introduction}

Magnetoelastic effects are ubiquitous in magnetic compounds. They may originate from modifications to the crystal field energies around the magnetic atoms. These modifications alter the contribution of the single ion anisotropy to the magnetic energy by changing the spin-orbit coupling~\cite{Kittel49}. An alternative mechanism, not requiring strong spin-orbit coupling, is exchange striction, that is a change of the strength of the effective exchange interactions between magnetic ions due to a modification of the interatomic bond lengths and angles~\cite{Tokura2014}. Thereby, atomic displacements in a crystal are susceptible to alter its magnetic order, as shown for instance in a variety of antiferromagnets \cite{Lines1965,Morosin1970,Lockwood1988}, spin spiral systems \cite{Kenzelmann2005,Sergienko2006}, frustrated magnets~\cite{Taniguchi2006,Heyer2006}. Exchange striction is additionally a source of lattice distortions and ferroelectricity in many of these materials. Such magnetic orders may arise simply due to a dominant antiferromagnetic (super)exchange interaction between neigbour spins, more subtly from a competition of ferro/antiferromagnetic interactions~\cite{Rastelli1979}, or also as a result of an antisymmetric Dzyaloshinski-Moriya exchange interaction~\cite{Cheong1989,Crepieux1998}.

In layered magnets, with the individual layers bond by van der Waals interactions (van der Waals magnets), exchange striction is also to be expected. The coupling between the spin and position degrees of freedom associated to the atoms (striction) in these materials acknowledgedly manifests in various ways: zone-folded phonons produced by the magnetic order~\cite{Scagliotti1987,Lee2016,Wang2016}, magnon-polarons~\cite{Vaclavkova2021,Liu2021,Pawbake2022} and thermal Hall effect \cite{Yang2024} in antiferromagnetic MPX$_3$ compounds, or a stiffening of phonon modes in CrBr$_3$ and CrSBr across magnetic phase transitions~\cite{Yin2021,Pawbake2023}, also observed in an isolated single-layer~\cite{Wu2022}. In metal oxychlorides and CrOCl in particular (see cartoon of the atomic structure in Figure~\ref{fig:exf}a), wherein a competition of antiferro/ferromagnetic spin interactions having comparable magnitude~\cite{Zhang2019b,Pawbake2025} spawns a rich phase diagram~\cite{Christensen1974,Angelkort2009,Reuvekamp2014,Zhang2014,Zhang2019,Zhang2023,Pawbake2025}, phonon stiffening and zone-folding have also been observed \cite{Zhang2019,Gu2022,Pawbake2025}. The multiple phases existing in various ranges of temperature and magnetic field are each liable to imprint a specific signatures in the atomic lattice. One of them, an antiferromagnetic state existing below 14~K, is associated to strong magnetoelectric effects, such that its magnetic state is efficiently controlled by an applied electric field in single/few-layer CrOCl devices~\cite{Gu2023}.

Here, we explore the succession of low-temperature phases that occur in CrOCl as a function of magnetic field, down to the single-layer (2D) limit, before the system becomes ferromagnetic at high fields, once the strong antiferromagnetic spin interactions have been overcome. Phases below 14~T have already been probed at this 2D limit, by tunneling magnetoconductance measurements that revealed an intermediate state between a low-field antiferromagnet and a higher-field ferrimagnet state, both commensurate to the atomic lattice~\cite{Zhang2023}. At high fields, yet other phases are known in bulk CrOCl~\cite{Pawbake2025}. Using magneto-Raman scattering spectroscopy at cryogenic temperature (5~K) up to 30~T (applied along the easy axis, perpendicular to the layers), we probe striction effects associated to all these phases, as function of the flakes' thickness. CrOCl is actually stable in air in the form of a single-layer, and its phonon modes are softer than in the multilayer case, as also observed in other van der Waals systems. We provide fine insights into the intermediate phase (between the antiferro/ferrimagnetic states), whose extension in field increases as thickness is reduced. Finally, we show that a canted ferrimagnetic phase exists down to the bilayer limit in the 20-30~T range, and that it responds to magnetic field with a strong quadratic stiffening of the phonon modes. The single-layer behaves distinctively, possibly with a (succession of) canted magnetic order(s).

\section{Results and discussion}

\subsection{Air-stable single-layer CrOCl}

\begin{figure}[!hbt]
    \includegraphics[width=80mm]{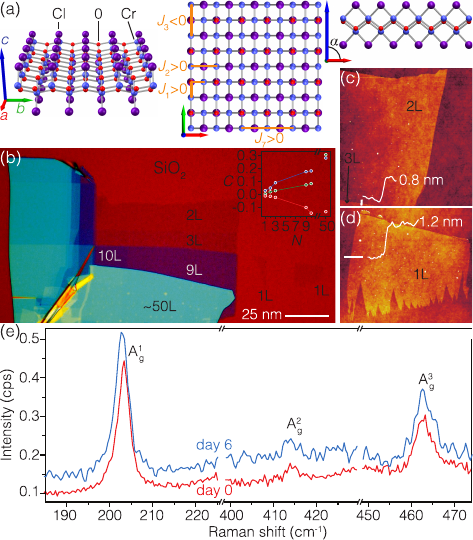}
    \caption{\label{fig:exf}\textbf{Air-stable 2D flakes of CrOCl exfoliated down to the monolayer.} (a) Ball-and-stick models of CrOCl's atomic structure (iso-structural to that of CrSBr), from different perspectives. The $\alpha$ angle goes off-90$^\circ$ in the orthorhombic phase. (b) Optical micrograph of a CrOCl flake exfoliated on SiO$_2$/Si. Inset: Optical contrast as function of the flake thickness (number of layers $N$). (c,d) AFM topographs of bilayer and monolayer regions, together with height profile across the step edges (along the white lines). (e) Raman scattering spectra (5~K, laser wavelength 515~nm) of a single layer, on the day of exfoliation and six days after.}
\end{figure}

Single crystals were synthesized by chemical vapour transport, see Ref.~\citenum{Pawbake2025}. They were pealed off using scotch tape and transfered directly onto SiO$_2$/Si substrates (subjected to an oxygen plasma beforehand). Few-layer flakes are identified by a dim optical contrast and with atomic force microscopy (Figure~\ref{fig:exf}b-d; see Methods section). On a staircase with varying number of layers, 0.8~nm step heights are measured between successive layers; the single-layer on the SiO$_2$ surface has a larger apparent height, of 1.2~nm (Figure~\ref{fig:exf}c), i.e. larger than the expected actual topography, presumably due to electrostatic forces~\cite{Arrighi2023}.

Raman scattering spectroscopy, performed here at 5~K (see Supplementary Information for room temperature data), reveals three vibrational modes that are characteristic of the symmetry of the crystal~\cite{Lee2021,Zhang2019}, the so-called $A_{\mathrm{g}}^1$, $A_{\mathrm{g}}^2$ and $A_{\mathrm{g}}^3$ phonon modes (Figure~\ref{fig:exf}e) consisting of out-of-plane vibrations, centered at $\sim$203~cm$^{-1}$, $\sim$414~cm$^{-1}$ and $\sim$462.5~cm$^{-1}$ respectively. The spectra do not change with time, at least at the scale of a week (Figure~\ref{fig:exf}e), indicating that material degradation upon air exposure is marginal or a slow process (note that the CrOCl flakes are not even capped with hexagonal boron nitride as was done in Refs.~\citenum{Gu2023,Zhang2023}). Relatively few truly 2D magnets are so stable, among which MnPS$_3$~\cite{Long2020}, NiPS$_3$~\cite{Lu2020}, CrPS$_4$~\cite{Son2021} and CrSBr~\cite{Klein2024}.

\subsection{Phonon softening with decreasing thickness}

\begin{figure}[!hbt]
    \includegraphics[width=80mm]{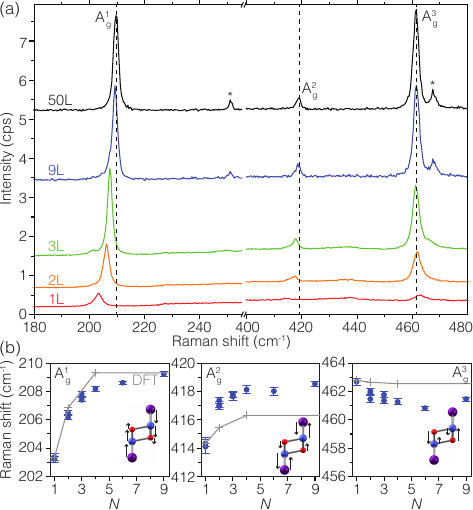}
    \caption{\label{fig:vibcoupling}\textbf{Interlayer vibrational coupling in few-layer CrOCl.} (a) Raman scattering spectra (5~K) of bulk and few-layer exfoliated CrOCl. The zone-folded phonon modes (labeled with a *) are related to the antiferromagnetic superstructure. (b) Frequency of the $A_{\mathrm{g}}^1$, $A_{\mathrm{g}}^2$ and $A_{\mathrm{g}}^3$ modes as function of the number of layers, from experiments (solid symbols) and from DFT calculations (gray crosses/lines; a scale factor of 5-6\% has been applied for wavenumber values to ease the comparison with the experimental data). Insets illustrate the atomic displacements associated with each phonon mode.}
\end{figure}

Figure~\ref{fig:vibcoupling}a presents the layer evolution of the phonon Raman scattering response. The $A_{\mathrm{g}}^1$, $A_{\mathrm{g}}^2$ and $A_{\mathrm{g}}^3$ phonons change energy with the number of layer $N$ with different trends --- a decrease (redshift) with $N$ for $A_{\mathrm{g}}^1$, $A_{\mathrm{g}}^2$ and a slight increase (blueshift) for $A_{\mathrm{g}}^3$ (Figure~\ref{fig:vibcoupling}b). This behaviour holds at room temperature too (see Supporting Information), and is reminiscent of observations made with the iso-structural compound CrSBr~\cite{Torres2023} (and van der Waals crystals in general).

This behaviour is rationalized by inspecting the nature of the three vibration modes (Figure~\ref{fig:vibcoupling}b, inset). The outer atoms in each single-layer (i.e. chlorine atoms) experience strong out-of-plane motion in the case of the $A_{\mathrm{g}}^1$ and $A_{\mathrm{g}}^2$ modes, but not in the $A_{\mathrm{g}}^3$ mode. This explains the former's larger susceptibility to the presence of neighbour layers, and their progressive softening with decreasing $N$ is an instance of the so-called `surface effect', wherein the outer atoms in each $N$-layer experience stronger force constants, reminiscent of observations made with MoS$_2$ \cite{Luo2013}. The effects seems the strongest with $A_{\mathrm{g}}^1$. For the $A_{\mathrm{g}}^3$ mode, no such effect is observed; on the contrary a slight stiffening is observed with decreasing $N$. The reason might relate to stacking-induced structural changes or the role of long-range Coulomb interactions \cite{Lee2010,Molina2011}. Density functional theory (DFT) calculations (see Methods sections), performed for the antiferromagnetic monoclinic phase of CrOCl relevant here (see below), reproduce, at least qualitatively, the observed trends.

\subsection{Evidence of magnetism in the bi- and single-layer}

\begin{figure}
    \includegraphics[width=80mm]{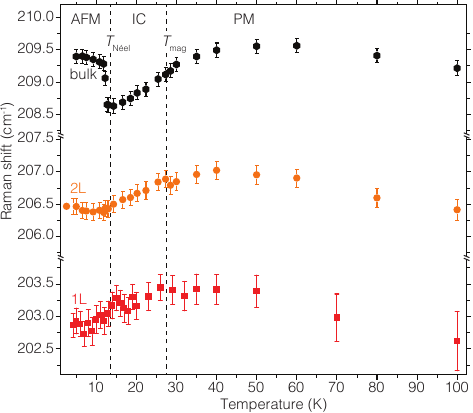}
    \caption{\label{fig:mag2D}\textbf{Inelastic scattering signature of 2D magnetism.} Frequency of the $A_{\mathrm{g}}^1$ phonon mode as function of temperature for bulk CrOCl and for a bilayer and a single-layer. The vertical dashed lines mark the N\'{e}el temperature $T_\mathrm{N\acute{e}el}$ to an antiferromagnetic (AFM) state and the magnetic incommensurate (IC) / paramagnet (PM) phase transition ($T_\mathrm{mag}$) in the bulk.}
\end{figure}

A first key question is whether CrOCl retains spontaneous magnetic order when thinned down to a few layers, and even to a single-layer. In its \textit{bulk} form, upon cooling down the material remains paramagnetic until $T_\mathrm{mag}\simeq$27~K, then adopts a magnetic order (unresolved so far to our knowledge) incommensurate with the atomic lattice, while the crystal still keeps the high temperature orthorhombic symmetry~\cite{Angelkort2009,Reuvekamp2014}. Further cooling down below $T_\mathrm{N\acute{e}el}$=14~K, CrOCl becomes antiferromagnetic with magnetic moments pointing perpendicular to the layers ($c$ axis), alternating up/down along the in-plane $a$ axis (forming parallel chains in the perpendicular $b$ axis) with a 1/4 commensurability with respect to the atomic lattice (Figure~\ref{fig:mag2D}, inset)~\cite{Christensen1974}. Concomitantly, the structure switches to a monoclinic one~\cite{Angelkort2009}, i.e. the angle $\beta$ between the $a$ and $c$ axes is about $90.07^\circ$ (Figure~\ref{fig:exf}a), and the $A_{\mathrm{g}}^1$ stiffens abruptly  (Figure~\ref{fig:mag2D}). This transition involves, not only a lateral sliding between successive layers (which are loosely bond by the van der Waals interaction), but also a shear within the $(a,c)$ plane \textit{in each single layer}.

Taking in mind the weak spin-orbit coupling effects in CrOCl~\cite{Wang2019,Qing2020}, the origin of these magneto-elastic phenomena is presumably not related to changes in the crystal-field experienced by the magnetic ions, but rather to an exchange striction mechanism. Considering the weak interlayer exchange (spin) interactions, typically two-to-three orders of magnitude smaller than intra-layer ones, it seems legitimate to wonder about the relative strength of the former and of the latter, and changes of bond length and angles might well have something to do with the appearance of magnetic order \textit{within} each layer.

Still in the bulk, a slow exchange-striction-related phonon softening is observed from $T_\mathrm{mag}$ to $T_\mathrm{N\acute{e}el}$ (Figure~\ref{fig:mag2D}).

Note that these striction effects yield a notable deviation from the expected progressive cooling-induced stiffening associated solely (i.e. in absence of magnetic orders) to anharmonic contributions to the interatomic potential energy (related to multi-phonon processes). The observations are altogether consistent with previous reports of similar behaviour for the $A_{\mathrm{g}}^2$ and $A_{\mathrm{g}}^3$ modes~\cite{Zhang2019,Gu2022}, and correlate with changes of the Raman modes' intensity~\cite{Yang2025}.

Surprisingly considering the weak exchange-like interaction between successive layers, exchange striction manifests differently in a bilayer or a single-layer than in the bulk  (Figure~\ref{fig:mag2D}), especially below 30~K. There either, the temperature-induced phonon anharmonicity alone cannot account for the observed progressive phonon softening, which is presumably related to exchange friction as in the incommensurate magnetic phase of the bulk. Below $\sim$14~K (i.e. close to the bulk $T_\mathrm{N\acute{e}el}$ value), the softening clearly stops for the bilayer, but it is unclear whether this also true for the single-layer. The data overall give strong indications of (a) spontaneous magnetic order(s) in both the bilayer and the single-layer below 25-30~K, yet the nature of these orders is unclear in the single-layer (based on the measurements we present here). We will now see that the magnetically ordered state in the bilayer, below 10-14~K, resembles that in the bulk.

\subsection{Intermediate phase(s) between low-field commensurate magnetic phases}

From now on, we turn to magnetic-field-controlled phases occurring at fixed temperature (5~K). The field is applied along the $c$ axis, the easy axis, and is increased from 0~T to 30~T. We first consider a range of moderate field values, from 0 to 10~T, also covered recently with tunneling magnetoconductance measurements on singe-/few-layers~\cite{Zhang2023}. In these previous works, the zero-field ground state has been identified, as being the same as in the bulk, regardless of the thickness. Also, the field-existence domain $[0~\mathrm{T},\sim 3.2~\mathrm{T}]$ of this state was found to be weakly dependent on the number of layers~\cite{Zhang2023}. Information about the higher-field phase, developing for field values larger than $\sim$3.8~T, is only available for bi-layers and thicker flakes. This phase is the ferrimagnetic one. Between the antiferromagnetic and ferrimagnetic phases, which are both commensurate with the atomic lattice, there are possibly two intermediate phases. The first one was proposed to appear after a spin-flop transition~\cite{Gu2022}. Whether it is distinct from a second phase, which might be the unresolved incommensurate phase living up to $T_\mathrm{mag}\simeq$~27~K in bulk~\cite{Zhang2023}, is unclear. We are interested here about the field domain of existence of this (these) intermediate phase(s), or more simply put, the field range delimited by the end of the antiferromagnetic phase and the beginning of the ferrimagnetic phase, and whether it depends on the flake thickness.

\begin{figure*}
    \includegraphics[width=158.06mm]{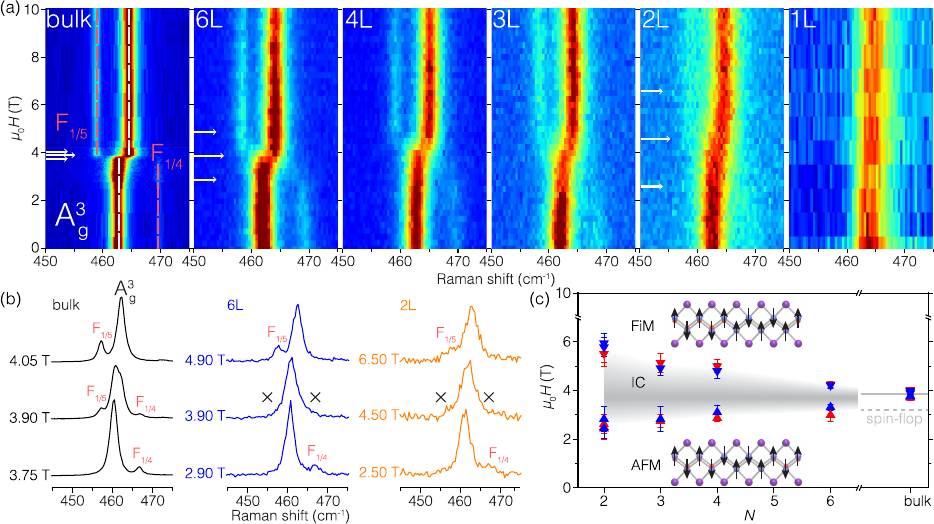}
    \caption{\label{fig:commensurate}\textbf{Spin-phonon coupling and zone-folding from a commensurate phase to the next one.} (a) Waterfall 2D maps of the Raman scattering spectra (5~K), around the $A_{\mathrm{g}}^3$ mode, acquired at increasing magnetic fields, for (from left to right) bulk CrOCl, six-, four-, three-, two- and single-layer flakes. (b) Selected spectra (arbitrary units) for the commensurate antiferromagnetic (1/4, low field) and ferrimagnetic (1/5, high field) phases, as well as for the intermediate field region, for the bulk (left), a six-layer (center) and a bilayer (right). (c) Field-domains of existence, derived from the Raman scattering data, of the 1/4-incommensurate (antiferromagnetic, AFM), incommensurate (IC) and 1/5-commensurate (ferrimagnetic, FiM) phases (0-10~T range), as a function of the number of layers. The spin-flop transition is indicated with a solid gray line. The spin arrangement for the two commensurate phases is sketched.}
\end{figure*}

In the bulk, the antiferromagnetic phase vanishes to the expense of the appearance of the ferrimagnetic phase within a very short magnetic field variation, as observed when tracking the intensity of their characteristic zone-folded phonons --- labelled $F_{1/4}$ and $F_{1/5}$ (in reference to the commensurability of the antiferro- and ferri-magnetic orders; Figure~\ref{fig:commensurate}a). In other words, the intermediate phases' domain of existence is very narrow in field, if it exists at all. Actually, within a narrow range of magnetic field of $\Delta B\sim50 mT$, we observe both $F_{1/4}$ and $F_{1/5}$ (Figure~\ref{fig:commensurate}b). The vanishing of the $F_{1/4}$ mode, and the appearance of the $F_{1/5}$, coincide with an abrupt shift of the nearby $A_{\mathrm{g}}^3$ mode frequency, as seen in Figure~\ref{fig:commensurate}a.

This magnetic-field-induced shift, also noticed for the bulk in Refs.~\citenum{Gu2022,Pawbake2025}, is the alter-ego of the temperature-induced one, observed at $T_\mathrm{N\acute{e}el}$, in the sense that it is also a signature of exchange striction, the magnetic order changing as the structure shears from the monoclinic system to the orthorhombic one according to density functional theory calculations~\cite{Gu2023}. The magnitude of the frequency shift relates to spin-spin correlations~\cite{Helman1968,Lockwood1988,Sushkov2005,Fennie2006} associated to each kind of order. It is however not trivial to infer the actual influence of these correlations since (\textit{i}) there are several exchange interactions at play in CrOCl and these interactions have different signs and involve not only the atomic positions of the Cr ions, but also those of the Cl and O atoms participating in the super-exchange paths~\cite{Zhang2019b,Pawbake2025}), and (\textit{ii}) their influence depends on the nature of the phonon modes $A_{\mathrm{g}}^{1,2,3}$, which feature atomic vibrations of different sorts.

When the flake thickness decreases, we observe that the field-region of the intermediate phase increases (Figure~\ref{fig:commensurate}a), extending across almost 3~T for the bilayer (Figure~\ref{fig:commensurate}c). Note that we observe no zone-folded phonons in this region (Figure~\ref{fig:commensurate}b), consistent with an assignment in terms of an incommensurate magnetic structure, which might hence be the very same phase as the incommensurate phase postulated at somehow higher temperature (but still below $T_\mathrm{mag}$)~\cite{Angelkort2009,Reuvekamp2014}. Interpreting how the incommensurate phase's field stability should depend on the thickness is not obvious. The origin may be slight changes of interlayer spin interactions, or of the structure associated to the presence of neighbour layers, indirectly altering super-exchange interactions (whose magnitude/sign depend sensitively on bond length/angle) or magnetic anisotropy.

\subsection{Canted magnetic phases}

\begin{figure*}
    \includegraphics[width=160mm]{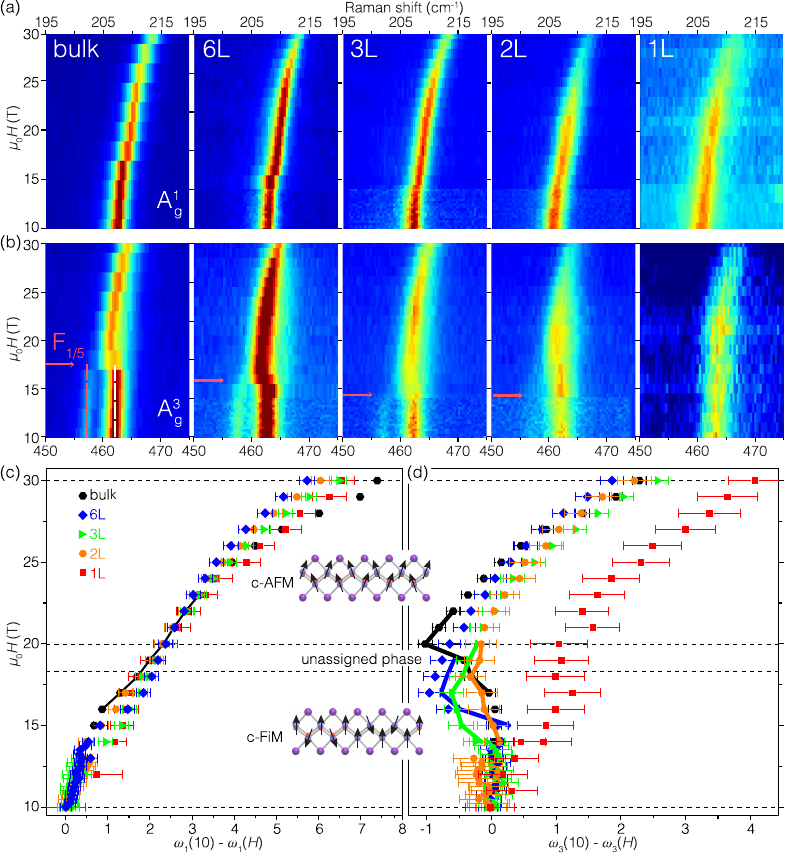}
    \caption{\label{fig:canted}\textbf{Spin-phonon coupling in a canted magnetic phase.} (a,b) Waterfall 2D maps of the Raman scattering spectra (5~K), around the $A_{\mathrm{g}}^1$ (a) and $A_{\mathrm{g}}^3$ (b) modes, in the 10-30~T range, for (left to right) 50- (i.e. bulk), six- , three-, two- and single-layer flakes. The red arrows mark the field values at which the $F_{1/5}$ folded mode vanishes. (c,d) Field-induced evolution of the frequency of the $A_{\mathrm{g}}^1$ ($\omega_1$) (c) and $A_{\mathrm{g}}^3$ ($\omega_3$) modes, with respect to their frequency at 10~T (d). Solid lines are quadratic fits to the data. In (d), thick lines connect the data points around the minimum of $\omega_3(10)-\omega_3(H)$. Inset: sketch of the canted ferrimagnetic (c-FiM) and antiferromagnetic (c-AFM) spin arrangements, with an intermediate, unassigned state in the intermediate field range. Note that the field was applied with a superconductive coil below 14~T and with a resistive one above.}
\end{figure*}

Around 10~T, in bulk CrOCl, a phase transition takes place from the ferrimagnetic state to a canted ferrimagnetic state (Figures~\ref{fig:commensurate}c,\ref{fig:canted}a), with the same 1/5 commensurability as evidenced by the persistence of the $F_{1/5}$ folded mode (Figure~\ref{fig:canted}a)~\cite{Pawbake2025}. Up to which precise field the canting angle of this phase increases is unclear in the literature so far. High-field magnetometry shows an obvious marked bump in the magnetization versus field derivative at 20~T \cite{Reuvekamp2014,Pawbake2025}, but careful inspection additionally reveals a small jump of the derivative before this value, at around 17~T (Supplementary Information). This appears to match with the disappearance of $F_{1/5}$ in our data (Figure~\ref{fig:canted}a). Concomitantly, the $A_{\mathrm{g}}^{3}$ mode starts to soften, up to 20~T, while all the way from 10~T up to 20~T, the $A_{\mathrm{g}}^{1}$ mode stiffens, with a change of slope around 17~T (Figure~\ref{fig:canted}a-d). The nature of the phase between 17~T and 20~T is unknown (we refer to it as `unassigned'); above 20~T, an antiferromagnetic canted phase with a 1/4 commensurability develops~\cite{Pawbake2025}.

Increasing the magnetic field from 20 to 30 T, the progressive shifts in the frequency of the $A_{\mathrm{g}}^{1,2,3}$ modes, of as much as 7~cm$^{-1}$ ($A_{\mathrm{g}}^{1}$) and 5~cm$^{-1}$ ($A_{\mathrm{g}}^{3}$) within 10~T, is yet another signature of exchange striction. Here, the changes of magnetic order are continuous, as the canting angle adjusts to mitigate the Zeeman energy; a maximum frequency shift corresponding to the perfect alignment of the spins along the field direction around 30~T, as the order becomes ferromagnetic~\cite{Pawbake2025}.

For few-layer flakes, the behaviour of the $A_{\mathrm{g}}^{1}$ mode is qualitatively very similar to that for the bulk: it monotonously stiffens with increasing field (Figure~\ref{fig:canted}a,c). The folded mode $F_{1/5}$ (not detected in the single-layer) disappears at lower fields, for smaller thicknesses than in the bulk (Figure~\ref{fig:canted}b), suggesting that the stability domain of the canted ferrimagnetic phase shrinks. Regarding the $A_{\mathrm{g}}^{3}$ mode, we observe, as for the bulk, first a plateau, then a softening and finally a stiffening, but the field value at the softening/stiffening tipping point is lower than the 20~T value for bulk (Figure~\ref{fig:canted}d), suggesting that the canted antiferromagnetic state appears at lower field.

The single-layer has a distinctive behaviour (also see Supporting Information for additional data), with no plateau in the $A_{\mathrm{g}}^{3}$ mode frequency, and, within uncertainties (which are larger than for thinner flakes due to the lower signal-to-noise ratio), a monotonous stiffening. The latter may correspond to a canted spin arrangement in the single-layer already at relatively low fields. We cannot however resolve the nature of these phases based on Raman scattering spectroscopy alone, nor can we determine whether they are commensurate (the corresponding folded modes would have low intensity presumably below noise level).

\begin{table}[!bt]
\caption{\label{tab:H2}High-field quadratic behaviour of the $A_{\mathrm{g}}^{1,3}$ modes' frequency with magnetic field $\omega^{(N)}_{1,3}(H)=\omega^{(N)}_{1,3}(H^{(N)}_{1,3})+\alpha^{(N)}_{1,3}(H-H^{(N)}_{1,3})^2$, for different number of layers $N$ in the flakes.}
\begin{tabular}{cccccc}
$N$ & $H^{(N)}_{1}$ & $\alpha^{(N)}_1$ & $H^{(N)}_{3}$ & $\alpha^{(N)}_3$ \\[0.1cm]
\hline
7 & 18.15 & 0.038 & 16.81 & 0.02 \\
6 & 16.11 & 0.018 & 15.87 & 0.013 \\
3 & 16.04 & 0.022 & 18.53 & 0.021 \\
2 & 13.76 & 0.016 & 18.3 & 0.017 \\
1 & 12 & 0.016 & 15.6 & 0.016 \\
\end{tabular}
\end{table}

The high-field evolution of the mode's frequency is well reproduced using a quadratic field function. The $\alpha^{(N)}_{1,3}$ pre-factor takes different values for the different modes, and changes with the number of layers $N$ in the flakes (see Table~\ref{tab:H2}). The $A_{g}^{2}$ phonon shows an evolution very similar with magnetic field but, because of its weaker intensity in the Raman scattering response, we present these results in the supplementary material.

\section{On the magnetic field dependence of the three high intensity modes}

To understand the experimentally observed quadratic dependence with the magnetic field of the $A_g^i$ modes, one can imagine that magneto-elastic interactions are at play.

\begin{figure}[htbp]
    \includegraphics[width=\columnwidth]{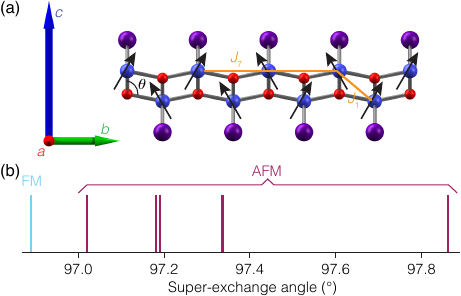}
    \caption{\label{fig:canted_magnetic order} a) Magnetic order of the canted 1$\times$4$\times$1 AFM phase of bulk CrOCl in the 20 to 30 T region. The Cr atoms are in blue and the O atoms in red. The exchange interactions $J_1$ and $J_7$ are the leading AFM interactions reported in reference \cite{Pawbake2025}. $\alpha$ is the super-exchange angle associated with the effective exchange interaction $J_1$. b) Comparison of the five super-exchange angles of the 1$\times$5$\times$1 FiM phase (blue) to the super-exchange angle of the FM phase (orange). }
\end{figure}

As already mentioned, the magnetic order, in the 20-30 T magnetic field region, is a canted 1$\times$4$\times$1 magnetic phase as illustrated in Fig.~\ref{fig:canted_magnetic order}a) and the inset of Fig.~\ref{fig:canted}.
In this canted phase, the projections of all the spins along the magnetic field contribute to the total magnetization while their projections along the \textit{b} axis compete with the magnetic field trying to satisfy the AFM exchange interactions $J_1 \simeq 9.9$~K and $J_7 \simeq 4.1$~K (see Fig.~\ref{fig:canted_magnetic order}a) and Ref.~\citenum{Pawbake2025}). The canting angle is a compromise between the contributions to the magnetic energy of the Zeeman and the exchange interaction terms.

When the magnetic field is increased, in the 20–30 T region, experiments show a linear increase of the magnetization~\cite{Pawbake2025}, corresponding to an increase in the projection of the spins along the direction of the magnetic field. As the system can no longer satisfy the leading AFM exchange interactions, it can \textit{limit the loss}, decreasing the AFM exchange interactions by diminishing the super-exchange angle $\alpha$ (see Fig.~\ref{fig:canted_magnetic order}a)). This effect is evident in our calculations, as illustrated in Fig.~\ref{fig:canted_magnetic order}b), where the five super-exchange angles of the 1$\times$5$\times$1 FiM phase are compared to the super-exchange angle of the FM phase.

The quadratic dependence with the magnetic field of the $A_g^i$ frequencies is a consequence of the above-mentioned magneto-elastic interaction. The energies (frequencies) of the three Raman active modes are the solutions of a standard eigenvalue problem:
\begin{equation}
  \mathbf{M}\sp{-1/2}\ \mathbf{D}\ \mathbf{M}\sp{-1/2}\ \vec{u}_i =
  \omega_i\sp2\  \vec{u}_i
\label{eq:eigen_phonons}
\end{equation}
where $\mathbf{M}$ is the mass matrix, $\mathbf{D}$ the Hessian of the potential energy, $\vec{u}_i = \mathbf{M}\sp{1/2}\ \vec{x}_i$, and $\vec{x}_i$ the vector with the atomic displacements.

Under an applied magnetic field, the magneto-elastic interactions slightly modify the atomic positions, changing the force constants in the Hessian matrix, and the phonon energies, which now depend on the magnetic field:
\begin{equation}
    \omega_i(H) = \sqrt{\frac{\vec{u}\sp{\intercal}_i\ \mathbf{M}\sp{-1/2}\ \mathbf{D}(H)\ \mathbf{M}\sp{-1/2}\ \vec{u}_i}
                    {\vec{u}\sp{\intercal}_i \cdot \vec{u}_i}}
\label{eq:frequency}
\end{equation}
Assuming that $\mathbf{D}(H) = \mathbf{D}(H_0) + \mathbf{\Delta D}$, we obtain at first order in $\mathbf{\Delta D}$:
\begin{equation}
    \omega_i(H) \simeq \omega_i(H_0) +
      \frac{\vec{u}\sp{\intercal}_i\ \mathbf{M}\sp{-1/2}\ \mathbf{\Delta D}\ \mathbf{M}\sp{-1/2}\ \vec{u}_i}
           {2\ \omega_i(H_0)\ \vec{u}\sp{\intercal}_i \cdot \vec{u}_i}
\label{eq:frequency}
\end{equation}
where $\vec{u}_i$, are the eigenvectors calculated with the magnetic field $H_0$.
One can use time reversal arguments to convince oneself that the effect can not depend on the sign of the magnetic field and the leading order must be quadratic:
\begin{equation}
    \mathbf{\Delta D}(H) =  \mathbf{A}\ (H-H_0)^2
\end{equation}
which results in
\begin{equation}
     \omega_i(H) =  \omega_i(H_0) + \alpha_i\ (H-H_0)^2
\end{equation}
where
\begin{equation}
     \alpha_i =  \frac{\vec{u}\sp{\intercal}_i\ \mathbf{M}\sp{-1/2}\ \mathbf{A}\ \mathbf{M}\sp{-1/2}\ \vec{u}_i}
                    {2\ \omega_i(H_0)\  \vec{u}\sp{\intercal}_i \cdot \vec{u}_i}
\end{equation}

Calculating the $\mathbf{A}$ matrix in the 20-30 T region requires evaluating the Hessian matrix for large cells, including spin-orbit interactions for various non-collinear magnetic orders. Unfortunately, this is beyond the scope of this paper. However, by comparing the Hessian matrices for the FiM and FM orders, we can estimate the order of magnitude of variation in the force constants:
\begin{equation}
    \frac{\partial^2 E}{\partial Cr_z \partial O_z}
\end{equation}
involved in the modification of the super-exchange angles discussed above (see Fig.~\ref{fig:canted_magnetic order}b), which are the important elements in the $\mathbf{A}$ matrix. Only the derivatives along $z$ are important, here, because the atomic displacements of the three $A_g^i$ modes are perpendicular to the planes.
The matrix elements are between -0.10 to -0.14 Ry/a$_0^2$ and their variation of the order of 0.005 Ry/a$_0^2$ which gives an energy shift of about 4 cm$^{-1}$ in good agreement with the experimental observations.

\section{Conclusion and outlook}

Chromium oxychloride is a van der Waals magnet exhibiting pronounced spin-lattice coupling. We explored the manifestation of exchange striction in the phonon spectrum of CrOCl, down to the single-layer thickness and as function of high magnetic fields, at cryogenic temperature. We find that the ground state, a commensurate antiferromagnet at zero field, transits to another commensurate phase, a ferrimagnet, through a spin flop transition and then an incommensurate phase. Above 10~T the magnetic order becomes canted (ferrimagnet) while remaining commensurate, and upon further increase of the field, adopts a yet unresolved magnetic phase before (starting from 20~T) reaching a canted antiferromagnetic state.

The phonon modes are softened as the thickness of the exfoliated flakes decreases, and are sensitive probes of the phase transitions. Complemented by the (dis)appearance of zone-folded phonon modes generated by the magnetic (in)commensurate superperiods, the softening or stiffening of the phonons is abrupt across transitions accompanied by a change of commensurability, and very progressive when a canting phase builds up magnetization under increased magnetic field. Strikingly, and although the interlayer magnetic interactions are very weak in CrOCl, the field-domains of existence of the different phases change quantitatively with the number of layers. While down to the bilayer, CrOCl seems to qualitatively behave as bulk CrOCl, the single-layer stands out, showing essentially smooth variations of the phonon modes' frequency, suggestive of canted phases above 3-4~T, and so far no sign of commensurability of the magnetic order.

Already very rich, the picture we build about multiple magnetic orders in CrOCl could be extended to resolve the nature of some of the phases we detected (around 3.5~T and 17~T in particular), to understand their temperature stability, and finally to understand the nature of magnetic order in the single-layer under a magnetic field. This will require to gain further information on the magnetic and structural degrees of freedom in CrOCl. The significant magnetoelastic effects together with the intrinsic competition of magnetic interactions suggest that the material should also be prone to a multitude of pressure/stress induced phases, up to very high fields (typically up to 30~T). Exploiting these phases in view of magnetoelectric functions, as done with some of them~\cite{Gu2023}, seems an exciting perspective.

\begin{acknowledgement}
This work was supported by LNCMI, member of the European Magnetic Field Laboratory (EMFL). Z.S. was supported by project LUAUS25268 from Ministry of Education Youth and Sports (MEYS) and by the project Advanced Functional Nanorobots (reg. No. CZ.$02.1.01/0.0/0.0/15 003/0000444$ financed by the EFRR). Z.S. acknowledge the assistance provided by the Advanced Multiscale Materials for Key Enabling Technologies project, supported by the Ministry of Education, Youth, and Sports of the Czech Republic and Project No. CZ.$02.01.01/00/22 008/0004558$, Co-funded by the European Union. This work was supported by the Agence Nationale de la Recherche (ANR) through project No. ANR-23-CE09-0034 `NEXT' and ANR-23-QUAC-0004. This work is supported by CEFIPRA project 7104-2. K.S acknowledges support from EMFL-ISABEL secondments program.

This work is also supported by France 2030 government investment plan 7 managed by the French National Research Agency under grant reference PEPR SPIN - SPINMAT ANR- 22-EXSP-0007.
\end{acknowledgement}

\bibliography{CrOCl_thin_layers}

\end{document}